**Sudoku-Inspired High-Shannon-Entropy Alloys**

Houlong Zhuang[†]

School for Engineering of Matter, Transport and Energy, Arizona State University, Tempe, AZ 85287, USA

[†]zhuanghl@asu.edu

**Abstract**

Current definition of high-entropy alloys (HEAs) is commonly based on the configurational entropy. But this definition depends only on the chemical composition of a HEA and therefore cannot distinguish the information content that is encoded in various local atomic arrangements and measurable by the Shannon entropy in information theory. Here, inspired by the finding that two-dimensional (2D) Sudoku matrices exhibit higher average Shannon entropy than 2D random matrix counterparts, we propose high-Shannon-entropy alloys (HSEA), whose structures are based on 3D Sudoku matrices. Despite the constraints on the atomic arrangements to form a 3D Sudoku $4 \times 4 \times 4$ matrix, we find that, a prototypical HSEA, NbMoTaW has a lower energy than the same HEA with a random structure due to the smaller lattice distortion. We compute the electronic structures and mechanical properties and find that the HSEA exhibits enhanced average Fermi velocity and ductility over the random NbMoTaW HEA. Finally, we evaluate the formation energies of single vacancies and a vacancy cluster in the HSEA, whose configurations mimic Sudoku puzzles with one and several missing numbers, respectively. We find that a range of large energies are required to generate such vacancies, which depend on the location and species of missing atoms. Our work shows that introducing the Shannon entropy to HEAs offers a useful metric to reveal more atomic details of HEAs and that HSEAs represent an uncharted domain in the complicated energy landscape of HEAs which may be endowed with improved properties.

**Introduction**

High-entropy alloys (HEAs) refer to alloys with five or more elements of equal or nearly equal concentration [1]. Located near or at the center of the high-dimensional composition space, HEAs are often associated with superior properties such as high corrosion resistance [2] and a good tradeoff between strength and ductility [3]. The term "high-entropy" in HEAs was originally coined to denote the configurational entropy, in the hope that a solid solution phase may be stabilized against competing intermetallic phases [4]. However, whether the stabilization of a solid solution is indeed mainly driven the contributions from the configurational entropy remains an open question [5]. Alternative appellations such as multi-principal element alloys (MPEAs) and complex concentrated alloys (CCAs) are therefore adopted to denote these compositionally disordered alloys. Regardless of the accuracy of its definition, HEA has an advantage over the alternatives, in the sense that the configurational entropy provides a straightforward metric to differentiate two alloys with different chemical compositions. For instance. The configurational entropy of the AlNbTaTiV HEA is higher than the $Al_{0.5}NbTaTiV$ HEA [6]. But this advantage diminishes when two HEAs have the same chemical composition or even the same HEA if two different simulation models are used. For example, two well-known HEAs, CoCrFeMnNi [7, 8] and AlCoCrFeNi [9, 10], nominally have the same configurational entropy but exhibit disparate information content at the atomic scale reflected by their dissimilar atomic structures subsequently manifested by distinct properties.

To recover this information content for a HEA, one can resort to information theory [11], founded by Claude E. Shannon [12], and invoke the concept of Shannon entropy defined therein. According to information theory, the amount of information of a random variable can be represented by "degree of surprise" on learning the variable [13]. The Shannon entropy embodies the average level of information in accord with the distribution of random variables. In the context of HEAs, the atomic arrangements of an HEA are random variables. In atomic-scale simulations, an HEA is modelled in many ways such as a special quasi-random structure (SQS) [14], a novel small set of ordered structures (SSOS) [15], or a random structure by permutating different species of the constituent elements in a given crystal lattice [16]. Computing the Shannon entropy therefore provides a route to distinguish these models, which can potentially lead to more accurate atomic details of an HEA.

To build a connection between the Shannon entropy and the atomic structure of an HEA, the species of the constituent elements can be assigned with different numerical identifiers (for example, 1, 2, 3, and 4 for a quaternary HEA), which can then be placed into slices of matrices to represent the atomic structure in detail. In the vast energy landscape of a HEA, it is possible that the identifiers in some slices form a 2D Sudoku matrix that undergirds the Sudoku puzzle.

The Sudoku puzzle is one of the most popular puzzle games often based on a 2D Sudoku $9 \times 9$ matrix that has $9 \times 9$ square grids distributed in $3 \times 3$ square blocks [17]. The matrix has a salient feature: each number from 1 to 9 must occur once and only once in each row, column, and square block. This feature unpins the rules of a Sudoku puzzle game. The game is to let the player identify the missing numbers a 2D Sudoku $9 \times 9$ matrix. Correspondingly, the difficulty level of a Sudoku puzzle can be adjusted by changing the number of missing numbers and their locations. Common variants of 2D Sudoku $9 \times 9$ matrices are 2D matrices of smaller matrix sizes such as $6 \times 6$ and $4 \times 4$. There are also 3D variants of Sudoku $9 \times 9 \times 9$ ($6 \times 6 \times 6$; $4 \times 4 \times 4$) matrices, where each of the 12 cross-sectional slice needs to be a 2D Sudoku $9 \times 9$ ($6 \times 6$; $4 \times 4$) matrix. Figure 1(a) illustrates a 3D Sudoku $4 \times 4 \times 4$ matrix with a visual aid for this matrix shown in Figure 1(b).

Once a full 2D Sudoku (e.g., 9 × 9) matrix is given, its Shannon entropy that measures the information content of the matrix can be computed. Because of the constraints on the locations of the numbers in a 2D Sudoku matrix, the Shannon entropy should intuitively be lower than a 2D random matrix. However, it has been shown that the average Shannon entropy of an ensemble of 2D Sudoku 9 × 9 matrices is significantly higher than that of the same number of random 2D 9 × 9 matrices [18]. This finding is counterintuitive, as one would expect that the more constraints required to form a 2D Sudoku 9 × 9 matrix leads to the less extent of disorder (in keeping with the less randomness) on the matrix and therefore the smaller Shannon entropy. Built upon this surprising finding, questions naturally arise as to (i) whether a 3D Sudoku 4 × 4 × 4 (we select 4 rather than 6 or 9 for the reason that quaternary HEAs are much more well studied) matrix encompassing 12 cross-sectional slices of 2D 4 × 4 matrices also has a larger average Shannon entropy than a random 3D 4 × 4 × 4 matrix? (ii) If the answer to (i) is positive, can the 3D Sudoku 4 × 4 × 4 matrix be mapped to form an atomic model for simulating an HEA with high Shannon entropy? (iii) What are the implications on the properties of the HEA?

To address these questions, we use the NbMoTaW HEA as an example to compute and compare the Shannon entropy and its concomitant properties, when the alloy is modelled using the structures based on a 3D Sudoku 4 × 4 × 4 matrix and a random matrix, respectively. This HEA is a prototypical body-centered cubic (BCC) refractory HEA that has been studied in many experiments [19-25] and theoretical calculations [26-31], and shown to exhibit excellent properties such as superior mechanical strength, promising for high-temperature engineering applications in aviation and nuclear energy [32]. We evaluate the relative stability of the two structures, examine their properties, and attempt to build a connection between the properties and the Shannon entropy. We focus on two critical properties of the NbMoTaW HEA, i.e., electrical conductivity and hardness. Both properties in thin film form of the HEA have been measured in two recent independent studies [22, 23].

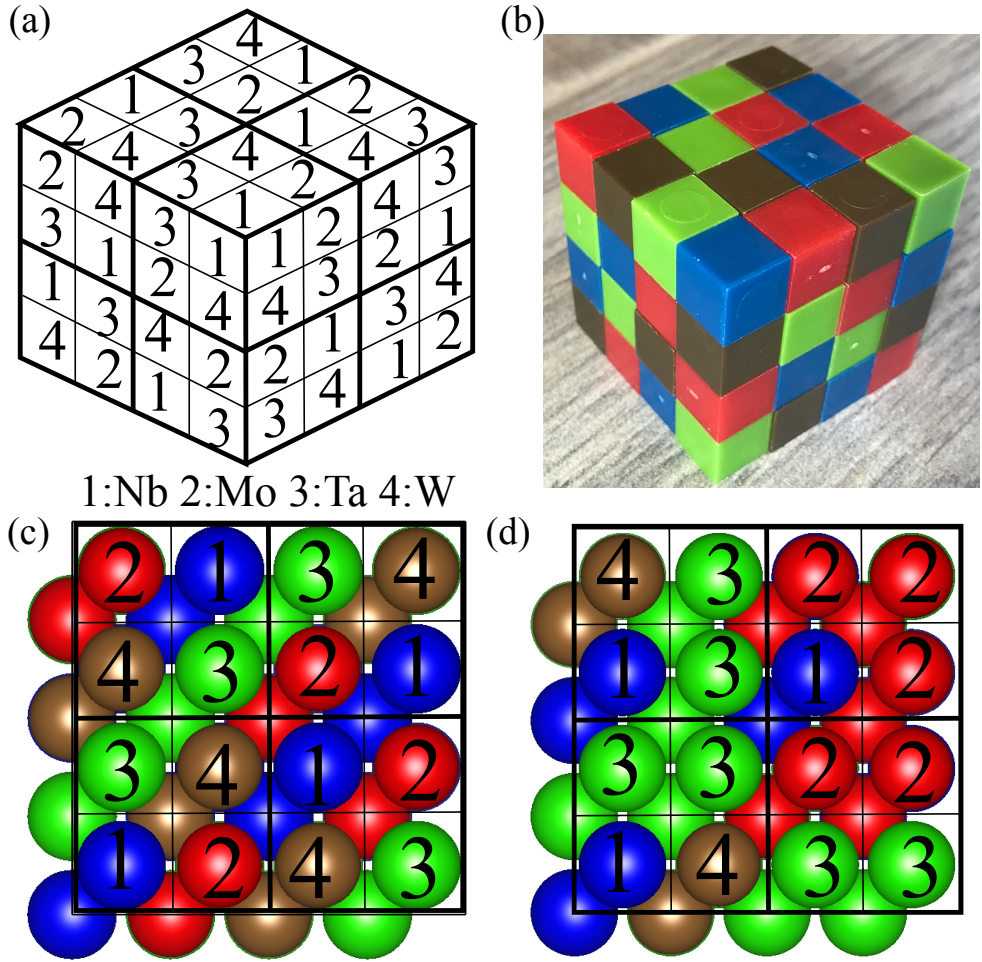

Figure 1. (a) An example of 3D Sudoku 4 × 4 × 4 matrix and (b) its corresponding visual aid using four colors of cubes. Top view of the atomic structures of the NbMoTaW high-entropy alloy that are based on (c) the 3D Sudoku matrix and on (d) a random 3D matrix.

**Simulation Methods**

We first create a mapping from a 3D Sudoku matrix (see Figure 1(a)) to the atomic structure of the NbMoTaW HEA. Each cross-sectional slice (viewed in the positive *a*, *b*, and *c* directions) of the 3D Sudoku consists of a 2D Sudoku 4 × 4 matrix. This 3D Sudoku matrix consists of 64 grid points, whose coordinates scaled by the sum of the experimental lattice constants of BCC Nb, Mo, Ta, and W can be used as the atomic coordinates in a cubic supercell. Without loss of generality, the numbers from 1 to 4 in the 3D Sudoku matrix can be assigned to Nb, Mo, Ta, and W, respectively. A single 3D Sudoku alone, however, can only be mapped to a simple cubic (SC) lattice. The BCC lattice of the NbMoTaW HEA is regarded as two interpenetrating SC lattices. For simplicity, we duplicate the existing SC lattice and then shift the duplicated lattice in the diagonal direction by a half of the diagonal length of a unit cell to form another SC lattice, which is followed by assigning atoms of the four species to the grids. One can of course use a different 3D Sudoku matrix and map it to the second sublattice or assign different IDs to the four elements. Doing so means exploring a different region of the energy landscape of the NbMoTaW HEA, which is in the purview of our future work. Still, the process to create the present model is sufficiently general and can be used to any other quaternary HEA with the BCC structure. For the purpose of comparison throughout this work, we remove the constraints on each cross-sectional slice and randomly assign numbers from 1 to 4 to each grid square. Following the same process, we obtain another BCC structure of NbMoTaW HEA with the sites randomly occupied by Nb, Mo, Ta, or W atoms. Each of the Sudoku-based or random structure consists of 128 atoms and has an equal amount (32) of the four types of atoms. Figure 1(c) and 1(d) shows the top views of the two atomic structures overlapped with 2D Sudoku and random matrices, respectively. We provide the Vienna *Ab initio* Simulation Package (VASP; version 5.4.4) input files in the POSCAR format in the Supporting Information.

We adopt the same approach of computing the Shannon entropy for 2D Sudoku matrices as used in Ref. [18]. Specifically, the process is as follows: First, we compute the singular values $\sigma_i$ via the singular value decomposition for the 12 cross-sectional 2D Sudoku 4 × 4 matrices. Next, for each of these matrices, we normalize the corresponding singular values (i.e., dividing any of the singular value by the sum of the four singular values) to obtain the normalized singular values $\hat{\sigma}_i$ that lie between 0 and 1. Analogous to the way of calculating the configurational entropy, the Shannon entropy $H$ is then calculated as $H = -\sum_{i=1}^{4} \hat{\sigma}_i \, log_2 \, \hat{\sigma}_i$. Here the logarithm to the base 2 is used to make the unit "bit". For an equal comparison, we compute the Shannon entropy of the 12 random 2D 4 × 4 matrices with overall the same counts of numbers from 1 to 4.

We use VASP for all the DFT calculations [33, 34], where the exchange-correlation interactions are approximated by the Perdew-Burke-Ernzerhof (PBE) [35] functional. To describe the electron-nuclei interactions, we use the PBE version of Nb, Mn, Ta, and W potential datasets generated via the projector augmented-wave (PAW) method [36, 37]. The $3s^2$ and $3p^2$ electrons of Nb atoms, $4s^2$ and $4p^2$ electrons of Mo atoms, $5s^2$ and $5p^2$ electrons of Ta atoms, and $5s^2$ and $5p^2$ electrons of W atoms are regarded as valence electrons in these PAW potentials. The kinetic-energy cutoff plane waves and the Monkhorst-Pack $k$-point grid [38] are set to 550 eV and 3 × 3 × 3, respectively. Before the calculations of electronic structures and mechanical properties, we fully optimize the 128-atom supercells using a force convergence criterion of 0.01 eV/Å. We use a denser $k$-point grid (6 × 6 × 6) for the calculations of the electronic structures including the Fermi surfaces, Fermi velocities, and density of states of the two structures.

To determine the mechanical properties, we calculate three independent elastic constants, $C_{11}$, $C_{12}$, and $C_{44}$ of the two nearly cubic structures. In principle, there should be 21 independent elastic constants for each system due to the low symmetry after geometry optimizations. We do not consider such an anisotropic effect that may occur in the elastic constant tensor of each structure. Namely, we assume that $C_{11}$, $C_{22}$, and $C_{33}$ are almost identical, so are $C_{12}$, $C_{13}$, and $C_{23}$, and are $C_{44}$, $C_{55}$, and $C_{66}$, and that the other 12 elastic constants are negligibly small. We argue that this is a reasonable assumption, as it has been shown that the 21 independent elastic constants of the face-centered cubic RhIrPdPtNiCu HEA have all been computed and the results are consistent with our assumption [39]. Following the strain-energy approach [40], we apply three sets of strains to the Sudoku or random ground-state supercells to obtain $C_{11}$, $C_{12}$, and $C_{44}$, respectively. For each set of applied strains (two positive and two negative strains), four strain energies in addition to the ground-state energy are used to fit to obtain the corresponding elastic constants. During the strain-energy calculations, we fix the shapes of the supercells and optimize only the atomic coordinates using the same force convergence criterion as above. We then calculate the bulk and shear moduli ($B_{VRH}$ and $G_{VRH}$, respectively) via the Voigt–Reuss–Hill (VRH) approximation [41-43]. We also compute Pugh's ratio, $K = G_{VRH}/B_{VRH}$ [44] and the Vickers hardness $H_V$ according to Chen's empirical formula [45]: $H_V = 2(K^2 G_{VRH})^{0.585} - 3$.

**Results and Discussion**

We first calculate the Shannon entropy of the two simulation systems. Figure 1 displays the computed Shannon entropy for all of the 12 2D Sudoku 4 × 4 matrices in each of the two systems. The four cross-sectional slices viewed in the *a*, *b*, and *c* directions are labeled with indices from 1 to 4, 5 to 8, and 9 to 12, respectively. For the Sudoku-based structure, most of the slice matrices have the identical Shannon entropy, showing that the permutations in the numbers have no effect on the Shannon entropy. By contrast, for the random structure, the distribution of Shannon entropies is more scattered, reflecting the random nature of the structure. We calculate the average entropy values of 12 matrices of each system and find that the Sudoku-based system has an average Shannon entropy of 1.459 bits whereas the random system's Shannon entropy is 1.313 bits. The higher Shannon entropy in the Sudoku-based system is consistent with the fact that a uniform probability distribution exhibits a larger amount of information content [13]. The Shannon entropy facilitates differentiating the two systems, which otherwise have the same configurational entropy. Due to the higher Shannon entropy caused by the embedded patterns of Sudoku matrices, we denote this system as high-Shannon-entropy alloy (HSEA) in contrast to the same HEA but with a random structure (henceforth referred to as RHEA).

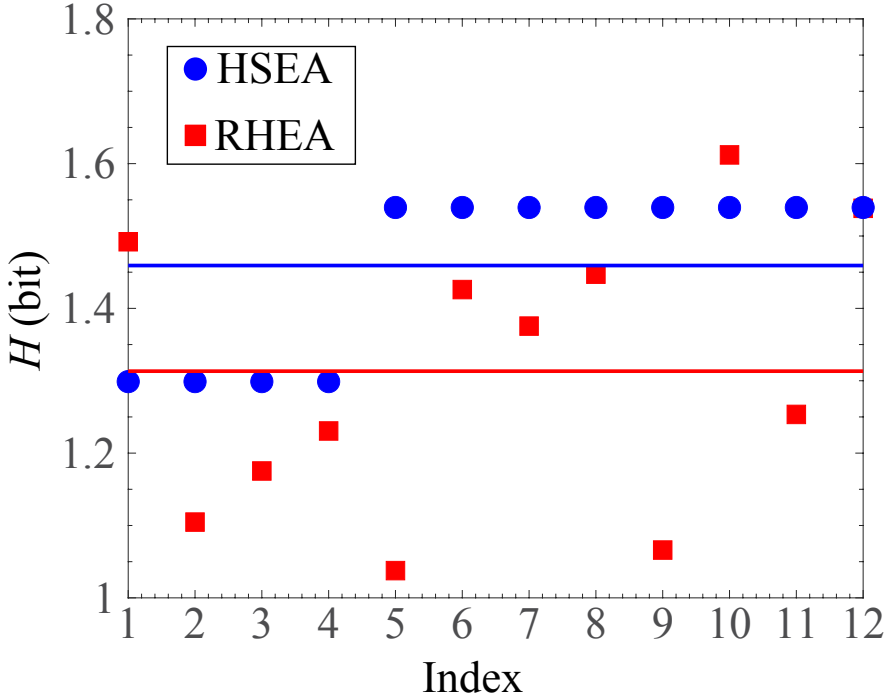

Figure 2. Shannon entropies ($H$) of the 12 cross-sectional slices (with indices from 1 to 12) of a 3D Sudoku 4 × 4 × 4 matrix and a 3D random matrix. Solid blue and red lines correspond to the average Shannon entropy of the Sudoku and random matrices, respectively. As described in the text, the atomic structure of a high-Shannon-entropy alloy (HSEA) is mapped from the Sudoku matrix and the same HEA with a random structure (RHEA) is from the random matrix.

Following the same argument that more constraints on Sudoku matrices correspond to a less structural disorder in the HSEA system, it may seem intuitive that the energy of the RHEA system should be lower, as we add more constraints to the atomic arrangement of atoms in a supercell. However, we compare the total energies of these two systems and find that they have similar total energies and the energy of HSEA is even slightly lower by 3.2 meV/atom. To understand why the energy of HSEA is smaller, we study the lattice distortion effect, which is also one of the four core effects of a general HEA [46]. We first compute the lattice parameters $a$, $b$, $c$, $\alpha$, $\beta$, and $\gamma$ listed in Table 1, which shows that the shapes of the supercells deviate from a cubic shape. Using these parameters alone nevertheless cannot provide complete information of the atomic locations in the supercells. The hard-sphere model is commonly used characterize the lattice distortion [47], but the lattice distortion from this model depends only on atomic radii and the concentrations of constituent elements, so the difference in the lattice distortions of the SHEA and RHEA supercells cannot be specified. We therefore propose to use the similarity between two vectors as a metric for the lattice distortion. The configurations for HSEA, RHEA, and an undistorted structure can be represented by three points in the 384 (128 × 3)-dimensional configurational space. The corresponding vectors are denoted as ($x_{1,x}$, $x_{1,y}$, $x_{1,z}$, $x_{2,x}$, $x_{2,y}$, $x_{2,z}$, … $x_{128,x}$, $x_{128,y}$, $x_{128,z}$), where $x_{i,x}$, $x_{i,y}$, $x_{i,z}$ are the three atomic coordinates for the $i^{th}$ atom. We then can compute the Euclidean distances between the HSEA (RSEA) configuration and the undistorted structure. The structure with the shorter Euclidean distance corresponds to the higher similarity to the undistorted structure and thus the smaller lattice distortion. We find that the similarity for the HSEA and RHEA configurations are 0.718 Å and 0.741 Å, respectively. We therefore conclude that RHEA exhibits a larger lattice distortion effect, causing the higher ground-state energy.

Table 1. Lattice parameters of a Sudoku-based NbMoTaW high-Shannon-entropy alloy (HSEA) (b) and HEA with a random structure (RHEA).

|      | $a$ (Å) | $b$ (Å) | $c$ (Å) | $\alpha$ (°) | $\beta$ (°) | $\gamma$ (°) |
|------|---------|---------|---------|--------------|-------------|--------------|
| HSEA | 12.942  | 12.941  | 12.933  | 89.845       | 89.839      | 89.675       |
| RHEA | 12.938  | 12.953  | 12.938  | 89.792       | 89.938      | 89.829       |

In terms of the properties of HSEA and RHEA, we begin with computing their electronic structures that play important roles in determining the electrical conductivities of the two systems. Here, we are not concerned about the exact numerical values of electrical conductivities from DFT calculations, which involve a number of factors that need to be taken into account, such as complex particle-particle interactions leading to different scattering mechanisms. Instead, we assume the applicability of the Drude free-electron model, so that the electrical conductivity is proportional to the Fermi velocity that describes the velocity of charge carriers and also to the density of electrons near the Fermi level that participate in electrical transport [48].

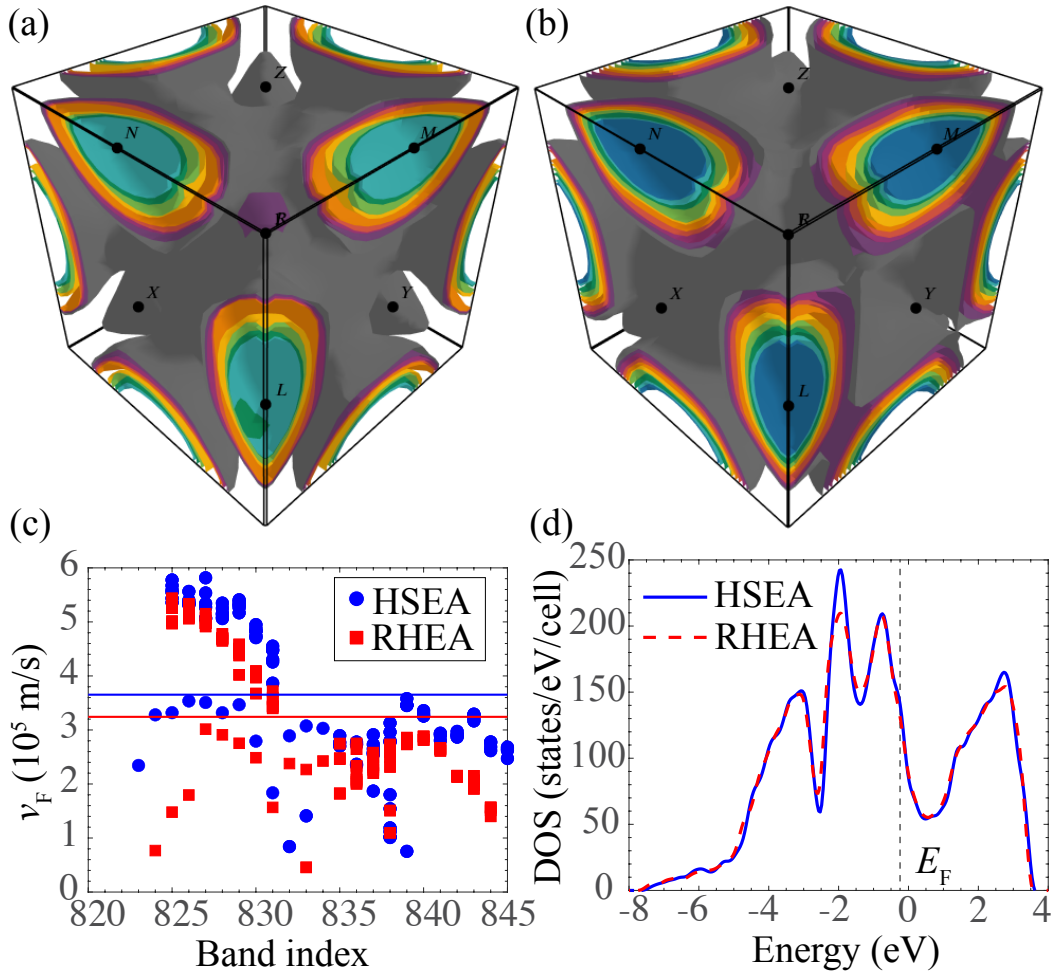

Figure 3. Fermi surfaces of (a) a Sudoku-based NbMoTaW high-Shannon-entropy alloy (HSEA) (b) and the same HEA with a random structure (RHEA). (c) Distributions of Fermi velocities $v_F$ and the corresponding average values (denoted by the solid blue and red lines) for all the Fermi-crossing bands in HSEA and RHEA. (d) Electron density of states (DOS) of the two systems.

Figure 3(a) and (b) illustrate the Fermi surfaces of HSEA and RHEA. We observe that there are 22 bands of the HSEA system crossing the Fermi level, whereas the number of bands crossing the Fermi level in the RHEA system is only one less. Furthermore, we obtain almost identical isosurfaces except that at the $Y$ point of the Fermi surface of RHEA appears in a rectangular shape, indicating electrons with the wavevectors behave anisotropically. Figure 3(c) summarizes the Fermi velocities of these bands, where we can see that the computed Fermi velocities of the two systems have similar distributions among the bands and that the RHEA system exhibits several bands having significantly smaller Fermi velocities. As a result, the average Fermi velocities of the HSEA and RHEA systems are $3.65 \times 10^5$ and $3.24 \times 10^5$ m/s, respectively. These values are much smaller than the computed average Fermi velocity of the constituent metals such as Mo and W with the Fermi velocities of $9.18 \times 10^5$ and $9.71 \times 10^5$ m/s, respectively [49]. The difference is not unexpected, as the lattice distortion reduces the electrical conductivity. A strong correlation has been found in many elemental metals, which shows that the higher Fermi velocity generally corresponds to the larger electrical conductivity given the same carrier density. We then calculate the electron density of states (DOS) shown in Figure 3(d), the two curves near their Fermi level are almost identical, implying that the number of electrons or charge carrier density that can be populated to the conduction bands and contribute to electrical transport is nearly the same. Combining the factors of the average Fermi velocity and DOS, we expect HSEA to exhibit a better electrical conductivity than RHEA.

We move on to calculate the mechanical properties of HSEA and RHEA. Table 2 shows the resulting elastic constants, bulk and shear moduli, Pugh's ratio and the Vickers hardness. We notice that all of these properties for the HSEA and RHEA systems are similar and the minor difference is reflected by the slightly smaller Pugh's ratio and Vickers hardness. Because Pugh's ratio is an indicator of ductility, the smaller Pugh's ratio of HSEA suggest the better ductility over RHEA. Furthermore, all the computed properties of HSEA and RHEA have no significant difference to those resulted from the SQS model used in Ref. [30], showing that the properties are not strongly dependent on the models used. In other words, both the HSEA and RHEA serve as appropriate models to compute the mechanical properties of an HEA. Note that all the three Vicker hardnesses shown in Table 2 are significantly smaller than the hardness measured for the NbMoTaW HEA in the thin film, for example, around 16.0 and 12.0 GPa in Refs. [22] and [23], respectively. This deviation is most likely due to the much smaller (at the nano scale) grain sizes causing the Hall-Petch effect in the thin-film system.

Table 2. Mechanical properties of a Sudoku-based NbMoTaW high-Shannon-entropy alloy (HSEA) and the same HEA with a random structure (RSEA), including three independent elastic constants $C_{11}$, $C_{12}$, and $C_{44}$, bulk modulus $B_{VRH}$ and shear modulus $G_{VRH}$, Pugh's ratio $K$, and the Vicker hardness $H_V$. The same properties of the same HEA modelled by a special quasi-random structure and resulted from Ref. [30] are shown for comparison.

|  | $C_{11}$ (GPa) | $C_{12}$ (GPa) | $C_{44}$ (GPa) | $B_{VRH}$ (GPa) | $G_{VRH}$ (GPa) | $K$ | $H_V$ (GPa) |
|---|---|---|---|---|---|---|---|
| HSEA | 370 | 160 | 61 | 230 | 76 | 0.33 | 3.89 |
| RHEA | 367 | 159 | 64 | 228 | 78 | 0.34 | 4.25 |
| Reference | 377 | 160 | 69 | 233 | 83 | 0.36 | 4.93 |

Thermodynamic vacancies are ubiquitous in alloys, so the NbMoTaW HSEA has no exception to possess such vacancies. Because the atomic structure of the HSEA is based on Sudoku matrices, the presence of vacancies is likened to missing numbers in the Sudoku matrices, which automatically gives rise to Sudoku puzzles. The simplest version of these Sudoku puzzles results from a single vacancy, corresponding to only one missing number in a 2D Sudoku 4 × 4 matrix and the solution to this type of puzzles is trivial. As the number of vacancies and their locations become complicated, the Sudoku puzzle becomes more challenging to solve. Figure 4(a) shows an example of the resulting Sudoku puzzle when a cluster of vacancies is generated on the first cross-sectional 2D Sudoku 4 × 4 matrix viewed along the positive *a* direction, i.e., the matrix with the index of 1 (see Figure 2). The solution to this Sudoku puzzle is provided in the Supporting Information.

We now proceed to evaluate the vacancy formation energies $E_f$ defined as: $E_f = E_{defect} + \frac{1}{128}(iE_{Nb} + jE_{Mo} + mE_{Ta} + nE_W) - E_{perfect}$, where $E_{defect}$ denotes the energy of the NbMoTaW HSEA supercell with a single vacancy (one of the integers $i$, $j$, $m$, or $n$ is equal to 1 and the others are equal to 0s) or a vacancy cluster ($i = j = m = n = 2$). $E_{Nb}$, $E_{Mo}$, $E_{Ta}$, and $E_W$ are the energies of fully optimized 128-atom 4 × 4 × 4 supercells of Nb, Mo, Ta, and W, respectively. $E_{perfect}$ is the total energy of the HSEA system without any vacancy. We focus on single vacancies and a vacancy cluster from the cross-sectional slice displayed in the Figure 4(a). Because the formation energy of a single vacancy is not unique, we calculate the formation energies of all the four possible single vacancies for each element, which results in the 16 different vacancy formation energies shown in Figure 4(c). Although the 16 Sudoku puzzles with a single missing number are equally straightforward to solve, the energies required to generate such single vacancies are different and strongly dependent on the locations and species of the missing atoms. For example, the $E_f$ of W single vacancies vary from 3.155 to 3.455 eV. Furthermore, although the $E_f$ of Mo single vacancies are generally higher than the those of other vacancies, when the Mo vacancy occurs at the third row of Figure 4(a), the $E_f$ (3.778 eV) is actually smaller than that of W single vacancy at the same row. The average $E_f$ values for the four constituent elements (3.411, 3.911, 3.370, and 3.801 eV for Nb, Mo, Ta, and W, respectively) agree well with the previous DFT study using an SQS cell (3.208, 4.078, 3.158, and 3.665 eV, respectively) [50] and also follow the same order: Ta < Nb < W < Mo. The relatively smaller average $E_f$ values of Ta and Nb single vacancies imply that they are more dominant vacancies at a fixed temperature. Furthermore, the calculated average $E_f$ are generally high than other quaternary HEAs such as CuNiCoFe (average $E_f$ = 2.147 eV) [51] and FeCoCrNi (average $E_f$ = 1.58-1.89 eV) [52], which suggests that the NbWTaW HSEA or HEA may exhibit high resistance to radiation damages. The formation energy of the vacancy cluster is 27.62 eV, which as expected is much larger than single vacancy formation energies and matches the drastic increase in the difficulty level of solving the Sudoku puzzle.

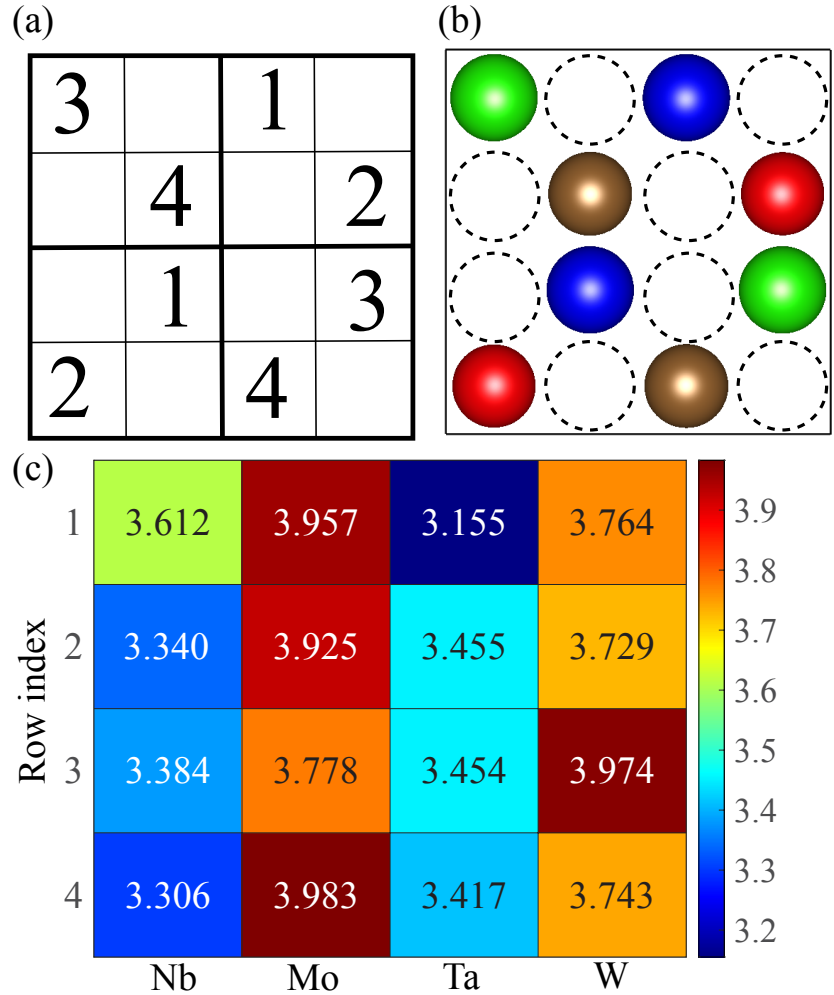

Figure 4. (a) A 2D Sudoku puzzle from a 2D Sudoku matrix (a cross-sectional slice taken from the 3D Sudoku matrix in Figure 1(a)) with missing numbers corresponding to (b) atomic vacancies of a cross-sectional slice of the unit cell of the Sudoku-based high-Shannon-entropy alloy. Dashed circles in (b) denote the locations of vacancies forming a vacancy cluster. (c) A heatmap of formation energies of Nb, Mo, Ta, and W single vacancies at the rows of (a).

**Conclusions**

Inspired by the Sudoku puzzle, we introduce the Shannon entropy to reveal more information content that is hidden in HEAs, so that two simulation supercells of a NbMoTaW HEA with the same configurational entropy can be differentiated by their magnitudes in the Shannon entropy. We find that the NbMoTaW HEA with the higher Shannon entropy is energetically more stable, electrically conductible, and mechanically malleable than the counterpart with the lower Shannon entropy. Finally, we study the formation energy of single vacancies and of a vacancy cluster in the HSEA, analogous to creating Sudoku puzzles with different number of missing numbers. We find that a wide range of formation energies for single vacancies that are on average higher than other quaternary HEAs, suggesting that the NbMoTaW HSEA is more resistant to radiation damages. We also show a large energy required to create a cluster of vacancies in the HSEA, which is consistent with the higher difficulty level of the corresponding Sudoku puzzle.

**Acknowledgements**

**References**

[1] D.B. Miracle, O.N. Senkov, A critical review of high entropy alloys and related concepts, Acta Materialia 122 (2017) 448-511.
[2] Y. Shi, B. Yang, P.K. Liaw, Corrosion-Resistant High-Entropy Alloys: A Review, Metals 7(2) (2017) 43.
[3] E. Ma, X. Wu, Tailoring heterogeneities in high-entropy alloys to promote strength–ductility synergy, Nature Communications 10(1) (2019) 5623.
[4] J.W. Yeh, S.K. Chen, S.J. Lin, J.Y. Gan, T.S. Chin, T.T. Shun, C.H. Tsau, S.Y. Chang, Nanostructured High-Entropy Alloys with Multiple Principal Elements: Novel Alloy Design Concepts and Outcomes, Advanced Engineering Materials 6(5) (2004) 299-303.
[5] A. Manzoor, S. Pandey, D. Chakraborty, S.R. Phillpot, D.S. Aidhy, Entropy contributions to phase stability in binary random solid solutions, npj Computational Materials 4(1) (2018) 47.
[6] X. Yang, Y. Zhang, P.K. Liaw, Microstructure and Compressive Properties of NbTiVTaAlx High Entropy Alloys, Procedia Engineering 36 (2012) 292-298.
[7] M.J. Yao, K.G. Pradeep, C.C. Tasan, D. Raabe, A novel, single phase, non-equiatomic FeMnNiCoCr high-entropy alloy with exceptional phase stability and tensile ductility, Scripta Materialia 72-73 (2014) 5-8.
[8] W.-M. Choi, Y.H. Jo, S.S. Sohn, S. Lee, B.-J. Lee, Understanding the physical metallurgy of the CoCrFeMnNi high-entropy alloy: an atomistic simulation study, npj Computational Materials 4(1) (2018) 1.

[9] Y. Zhang, T.T. Zuo, Z. Tang, M.C. Gao, K.A. Dahmen, P.K. Liaw, Z.P. Lu, Microstructures and properties of high-entropy alloys, Progress in Materials Science 61 (2014) 1-93.
[10] K. Yamanaka, H. Shiratori, M. Mori, K. Omura, T. Fujieda, K. Kuwabara, A. Chiba, Corrosion mechanism of an equimolar AlCoCrFeNi high-entropy alloy additively manufactured by electron beam melting, npj Materials Degradation 4(1) (2020) 24.
[11] T.M. Cover, Elements of information theory, John Wiley & Sons1999.
[12] G.P. Collins, Claude E. Shannon: Founder of Information Theory, Scientific American 14 (2002).
[13] C.M. Bishop, Pattern recognition, Machine learning 128(9) (2006).
[14] A. Zunger, S.H. Wei, L.G. Ferreira, J.E. Bernard, Special quasirandom structures, Physical Review Letters 65(3) (1990) 353-356.
[15] C. Jiang, B.P. Uberuaga, Efficient Ab initio Modeling of Random Multicomponent Alloys, Physical Review Letters 116(10) (2016) 105501.
[16] D. Wang, L. Liu, M. Chen, H. Zhuang, Electrical and thermal transport properties of medium-entropy $Si_yGe_ySn_x$ alloys, Acta Materialia 199 (2020) 443-452.
[17] J.-P. Delahaye, The science behind Sudoku, Scientific American 294(6) (2006) 80-87.
[18] P.K. Newton, S.A. DeSalvo, The Shannon entropy of Sudoku matrices, Proceedings of the Royal Society A: Mathematical, Physical and Engineering Sciences 466(2119) (2010) 1957-1975.
[19] O.N. Senkov, G.B. Wilks, D.B. Miracle, C.P. Chuang, P.K. Liaw, Refractory high-entropy alloys, Intermetallics 18(9) (2010) 1758-1765.
[20] O.N. Senkov, G.B. Wilks, J.M. Scott, D.B. Miracle, Mechanical properties of $Nb_{25}Mo_{25}Ta_{25}W_{25}$ and $V_{20}Nb_{20}Mo_{20}Ta_{20}W_{20}$ refractory high entropy alloys, Intermetallics 19(5) (2011) 698-706.
[21] Y. Zou, H. Ma, R. Spolenak, Ultrastrong ductile and stable high-entropy alloys at small scales, Nature Communications 6(1) (2015) 7748.
[22] X. Feng, J. Zhang, Z. Xia, W. Fu, K. Wu, G. Liu, J. Sun, Stable nanocrystalline NbMoTaW high entropy alloy thin films with excellent mechanical and electrical properties, Materials Letters 210 (2018) 84-87.
[23] H. Kim, S. Nam, A. Roh, M. Son, M.-H. Ham, J.-H. Kim, H. Choi, Mechanical and electrical properties of NbMoTaW refractory high-entropy alloy thin films, International Journal of Refractory Metals and Hard Materials 80 (2019) 286-291.
[24] X. Feng, J. Utama Surjadi, Y. Lu, Annealing-induced abnormal hardening in nanocrystalline NbMoTaW high-entropy alloy thin films, Materials Letters 275 (2020) 128097.
[25] Y. Zou, S. Maiti, W. Steurer, R. Spolenak, Size-dependent plasticity in an $Nb_{25}Mo_{25}Ta_{25}W_{25}$ refractory high-entropy alloy, Acta Materialia 65 (2014) 85-97.
[26] W.P. Huhn, M. Widom, Prediction of A2 to B2 Phase Transition in the High-Entropy Alloy Mo-Nb-Ta-W, JOM 65(12) (2013) 1772-1779.
[27] F. Körmann, M.H.F. Sluiter, Interplay between Lattice Distortions, Vibrations and Phase Stability in NbMoTaW High Entropy Alloys, Entropy 18(8) (2016) 403.
[28] F. Körmann, A.V. Ruban, M.H.F. Sluiter, Long-ranged interactions in bcc NbMoTaW high-entropy alloys, Materials Research Letters 5(1) (2017) 35-40.
[29] Y.L. Hu, L.H. Bai, Y.G. Tong, D.Y. Deng, X.B. Liang, J. Zhang, Y.J. Li, Y.X. Chen, First-principle calculation investigation of NbMoTaW based refractory high entropy alloys, Journal of Alloys and Compounds 827 (2020) 153963.

[30] X.-G. Li, C. Chen, H. Zheng, Y. Zuo, S.P. Ong, Complex strengthening mechanisms in the NbMoTaW multi-principal element alloy, npj Computational Materials 6(1) (2020) 70.
[31] S. Mu, S. Wimmer, S. Mankovsky, H. Ebert, G.M. Stocks, Influence of local lattice distortions on electrical transport of refractory high entropy alloys, Scripta Materialia 170 (2019) 189-194.
[32] Z.D. Han, H.W. Luan, X. Liu, N. Chen, X.Y. Li, Y. Shao, K.F. Yao, Microstructures and mechanical properties of TixNbMoTaW refractory high-entropy alloys, Materials Science and Engineering: A 712 (2018) 380-385.
[33] P. Hohenberg, W. Kohn, Inhomogeneous electron gas, Physical Review 136(3B) (1964) B864.
[34] W. Kohn, L.J. Sham, Self-consistent equations including exchange and correlation effects, Physical Review 140(4A) (1965) A1133.
[35] Perdew, Burke, Ernzerhof, Generalized gradient approximation made simple, Physical Review Letters 77(18) (1996) 3865-3868.
[36] P.E. Blöchl, Projector augmented-wave method, Physical Review B 50(24) (1994) 17953-17979.
[37] G. Kresse, D. Joubert, From ultrasoft pseudopotentials to the projector augmented-wave method, Physical Review B 59(3) (1999) 1758-1775.
[38] H.J. Monkhorst, J.D. Pack, Special points for Brillouin-zone integrations, Physical Review B 13(12) (1976) 5188.
[39] B. Yin, W.A. Curtin, First-principles-based prediction of yield strength in the RhIrPdPtNiCu high-entropy alloy, npj Computational Materials 5(1) (2019) 14.
[40] H. Zhuang, M. Chen, E.A. Carter, Elastic and Thermodynamic Properties of Complex Mg-Al Intermetallic Compounds via Orbital-Free Density Functional Theory, Physical Review Applied 5(6) (2016) 064021.
[41] W. Voight, Lehrbuch der kristallphysik, Teubner, Leipzig (1928).
[42] A. Reuß, Berechnung der fließgrenze von mischkristallen auf grund der plastizitätsbedingung für einkristalle, ZAMM‐Journal of Applied Mathematics and Mechanics/Zeitschrift für Angewandte Mathematik und Mechanik 9(1) (1929) 49-58.
[43] R. Hill, The elastic behaviour of a crystalline aggregate, Proceedings of the Physical Society. Section A 65(5) (1952) 349.
[44] S.F. Pugh, XCII. Relations between the elastic moduli and the plastic properties of polycrystalline pure metals, The London, Edinburgh, and Dublin Philosophical Magazine and Journal of Science 45(367) (1954) 823-843.
[45] X.-Q. Chen, H. Niu, D. Li, Y. Li, Modeling hardness of polycrystalline materials and bulk metallic glasses, Intermetallics 19(9) (2011) 1275-1281.
[46] J.-W. Yeh, Physical Metallurgy of High-Entropy Alloys, JOM 67(10) (2015) 2254-2261.
[47] Q. He, Y. Yang, On Lattice Distortion in High Entropy Alloys, Frontiers in Materials 5(42) (2018).
[48] N.W. Ashcroft, N.D. Mermin, Solid state physics, holt, rinehart and winston, new york London, 1976.
[49] D. Gall, Electron mean free path in elemental metals, Journal of Applied Physics 119(8) (2016) 085101.
[50] L.D. Nibbelink, Simulating Vacancy Formation and Diffusion in NbMoTaW, University of California, San Diego2020.

[51] A. Esfandiarpour, M.N. Nasrabadi, Vacancy formation energy in CuNiCo equimolar alloy and CuNiCoFe high entropy alloy: ab initio based study, Calphad 66 (2019) 101634.
[52] W. Chen, X. Ding, Y. Feng, X. Liu, K. Liu, Z.P. Lu, D. Li, Y. Li, C.T. Liu, X.-Q. Chen, Vacancy formation enthalpies of high-entropy FeCoCrNi alloy via first-principles calculations and possible implications to its superior radiation tolerance, Journal of Materials Science & Technology 34(2) (2018) 355-364.